\documentclass[aps,prb,twocolumn,groupedaddress,floats,showpacs]{revtex4}
\usepackage{latexsym}
\usepackage{dcolumn}
\usepackage[dvips]{graphicx}
\usepackage{amssymb}
\usepackage{graphics}
\usepackage{amsmath}
\usepackage{epsf}

\newcommand{\vk}{{\bf k}}

\begin{document}
\author{E. H. Hwang and S. Das Sarma}
\title{Limit to 2D mobility in modulation-doped GaAs quantum structures: How
  to achieve a mobility of 100 millions}
\affiliation{Condensed Matter Theory Center, Department of Physics, 
University of Maryland, College Park, MD 20742-4111}
\date{\today}
\begin{abstract}
Considering scattering by unintentional background charged impurities
and by charged dopants in the modulation doping layer as well as by
GaAs acoustic phonons, we theoretically consider the practical
intrinsic (phonons) and extrinsic (background and dopants) limits to
carrier mobility in modulation doped AlGaAs-GaAs 2D semiconductor
structures. We find that reducing background impurity density to
$10^{12}$ cm$^{-3}$ along with a modulation doping separation of 1000 \AA
\; or above will achieve a mobility of $100 \times 10^6$ cm$^2$/Vs at a carrier
density of $3\times 10^{11}$ cm$^{-2}$ for $T=1$K. At $T=4$ (10)K,
however, the hard limit to the 2D mobility
would be set by acoustic phonon scattering with the maximum
intrinsic mobility being no higher than 22 $(5) \times 10^6$
cm$^2$/Vs. Detailed numerical results are presented as a function of
carrier density, modulation doping distance, and temperature to
provide a quantitative guide to experimental efforts for achieving
ultra-high 2D mobilities. 
\end{abstract}

\maketitle

\vspace{0.5cm}

%%%%%%%%%%%%%%%%%%%%%%%%%%%%%%%%%%%%%%%%%%%%%%%%%%%%%%%%%%%%%%%%

One of the most spectacular achievements of modern materials science
is the continued enhancement of low temperature ($T \alt 1$K) 2D electron
mobilities in confined GaAs structures (e.g. quantum wells,
heterostructures) over the last 30 years
\cite{stormer,pfeiffer,heiblum}. In particular, the invention 
of modulation doping technique \cite{stormer} allowing spatial
separation between 
dopants and carriers in 2D semiconductor quantum structures and the
continuous improvement in molecular beam epitaxy (MBE) 
eliminating unintentional charged impurities from the
background have led to an exponential increase in the low-temperature
2D carrier mobility in modulation doped GaAs-AlGaAs quantum
structures
%but there have been notable mobility enhancement in other
%2D structures also such as undoped GaAs HIGFET \cite{Kane} and
%modulation doped Si-SiGe systems\cite{sige})
from roughly $5\times 10^3$ cm$^2$/Vs in
1977\cite{stormer} to
$3.6\times 10^7$ cm$^2$/Vs in 2007\cite{heiblum}, an amazing increase of
mobility by a 
factor of 7000 in 30 years. Although much less widely known than
Moore's law, this is roughly equivalent to a factor of 3 increase in
effective carrier velocity  for every three years, more
or less the same exponential rate of increase as in the celebrated
Moore's law for Si microprocessors except that instead of
miniaturization, which drives the Moore's law in Si electronics,
modulation doping directly enhances the low temperature carrier
mobility. This astonishing 7000-fold low-temperature mobility
enhancement in GaAs based 2D systems has been possible entirely
through the systematic suppression of charged impurity scattering and
interface roughness scattering in GaAs-AlGaAs heterostructures and
AlGaAs-GaAs-AlGaAs quantum wells (which can be thought of as double
heterostructures). This exponential increases in 2D mobility is
therefore a triumph of materials science, {\it not} lithographic
processing. 
The primary motivation for enhancing 2D mobility comes from the
physics of fractional quantum Hall effect (FQHE) where higher mobility
leads invariably to the discovery of new phenomena \cite{Dean}. Recent
theoretical developments \cite{dassarma} on topological quantum
computation have 
provided added impetus \cite{heiblum} to increasing 2D mobilities as
high as possible.

While improvement in materials science and device fabrication has
continued to enhance the low temperature ($\alt 1$K)
disorder-scattering limited 2D  mobility, the question naturally
arises on the possible {\it intrinsic}, rather than the {\it
  extrinsic} (i.e. limited by background or remote charged impurities,
interface roughness, etc.) in quantum 2D structures. This intrinsic
limit, which must depend on the temperature, is obviously determined
by phonon scattering since phonons, unlike impurities, can never be
eliminated from a system except at $T=0$K.
This intrinsic temperature dependent mobility limit
imposed by phonon scattering is sometimes referred to as the {\it
  hard} mobility limit in contrast to the {\it soft} limit imposed by
impurity scattering or interface roughness scattering or alloy
scattering which can, in principle, be reduced or even eliminated
through materials science advances. The only way to reduce the phonon
scattering is to reduce the temperature, and at any particular
temperature, the phonon scattering indeed sets an impassable hard
intrinsic limit to the mobility at that temperature.

In addition to the obvious contributions by phonons to the intrinsic
resistivity, there is actually an additional intrinsic contribution
arising from the scattering by the charged dopants associated with the
modulation doping producing the carriers themselves. This is a rather
subtle effect that arises in semiconductors because an {\it intrinsic}
semiconductor does not have free charge carriers (as, for example,
metals do with the metallic Fermi level lying in the conduction band),
and any mobile electron in a semiconductor must leave behind a
positively charged dopant somewhere.  The scattering between the
dopant ions and the carriers must be considered an intrinsic
resistive mechanism since a complete removal of the dopants would also
eliminate the carriers themselves. Of course, the whole point or
motivation for the modulation doping technique is to strongly suppress
carrier scattering by the dopant ions through the large spatial
separation between the carriers in the 2DEG in GaAs and the dopants
placed in a thin layer (so-called delta-layer) in AlGaAs at a distance $d$
from the 2D carriers. Making the simple assumption, which is 
accurate in this case, that the dopant ions and the 2D electron gas
form together a parallel plate capacitor, we get ($e$ is the electron
charge):
$n=\kappa |\Delta V|/(4\pi|e|d)$,
where $n$, $\kappa$, $d$, and $|V|$ are the 2D carrier density, the
background lattice dielectric constant of the insulating regime
(i.e. Ga$_{1-x}$Al$_x$As), the modulation doping separation between
the delta doping layer in AlGaAs and the 2DEG in GaAs, and the voltage
or the energy barrier at the GaAs-Ga$_{1-x}$Al$_x$As interface (which
depends on the Al molar fraction $x$) which creates the confining
potential producing the 2DEG. We note that the maximum possible value
of $|\Delta V| \approx 0.8$ eV for $x=0$, and for a given $x$, the
value of the modulation separation $d$ determines the carrier
density. In our discussion of intrinsic mobility we will ignore the
precise relationship between $d$ and $n$, and assume that $d$ and
$n$ are independent variables as far as the 2D mobility calculation
goes. This is mathematically justified because $|\Delta V|$ could, in
principle, be adjusted, making $d$ and $n$ independent variables.

The 2D conductivity $\sigma$ and mobility $\mu$ are given by
$\sigma=ne\mu=ne^2\tau/m$ where $\tau$ is the scattering time or the
transport relaxation time. There are many independent contributions to
the scattering rate $\tau^{-1}$ in the modulation-doped GaAs-AlGaAs
2DEG system: (1) Acoustic phonon scattering via deformation coupling;
(2) acoustic phonon scattering via piezoelectric coupling; 
(3) impurity scattering by unintentional
background charged impurities invariably present in the background;
(4) impurity scattering by intentional dopants in the modulation-doping
delta layer; (5) interface roughness scattering at the GaAs-AlGaAs
interfaces; (6) alloy disorder scattering arising in
Ga$_x$Al$_{1-x}$As; 
(7) LO-phonon scattering via long-range polar Fr\"ohlich coupling; 
(8) short-range scattering by neutral defects and impurities. Assuming
Matthiessen's rule, which is strictly valid only at $T=0$ (but is
reasonably valid for our purpose at higher temperatures also), we can
write $\tau^{-1} = \sum_i \tau_i^{-1}$, where $\tau_i$ indicates the
transport relaxation time due to various individual scattering
mechanisms. As discussed above, only the first two mechanism arising
from the acoustic phonon scattering are true intrinsic resistive
mechanism, but we will also include in our consideration the Coulomb
scattering due to the intentional dopants in the modulation doping
layer as well as  the scattering by the unintentional random charged
impurities invariably present in the background since in currently
existing ultra-high mobility 2D samples, the background random charged
impurity scattering is known to be the mobility-limiting mechanism.
We note that the other scattering mechanisms (items 5 to 8 listed
above, are known to be much less quantitatively important than the
mechanisms (items 1 to 4) we are considering in this work.
Polar carrier scattering by LO phonons is important in GaAs only at
relatively high temperatures ($>50$K), becoming dominant at room
temperatures.  Due to the rather high energy of the GaAs optical
phonons ($\sim 36$ meV), LO phonon scattering is completely suppressed in
the temperature regime ($<10$K) of interest to us in this work.
Resistive scattering by optical phonons in 2D 
GaAs system has been considered in the literature \cite{kawamura2}.

We start by considering the Coulomb scattering by the 2D remote
charged dopants in the modulation doping layer which must invariably
be present in modulation doped GaAs-AlGaAs structures and by the
unintentional 3D background charged impurities. Using the Boltzmann
theory, the scattering rate, $\tau^{-1}_c$, due to random charged
Coulomb impurities, either in the background (i.e. 3D unintentional
dopants) or in the modulation doping layer (i.e. in a 2D layer a
distance $d$ from the 2DEG), is given by \cite{dassarma1,RMP}
\begin{eqnarray}
\frac{1}{\tau_c(\varepsilon_{\vk})} & = & \frac{2\pi}{\hbar}\sum_{\alpha}
\int \frac{d^2k'}{(2\pi)^2}  
\int^{\infty}_{-\infty}dzN_i^{(\alpha)}(z) \nonumber \\
& \times & |u_{ci}^{(\alpha)}({\bf k}-{\bf k}';z)|^2 
(1-\cos \theta_{{\bf k k}'})
\delta(\varepsilon_{\vk}-\varepsilon_{\vk'}), 
\label{itau}
\end{eqnarray}
where $N_i^{(\alpha)}(z)$ is the 3D distribution of the $\alpha$-th
kind of charged impurities, $u_{ci}^{(\alpha)}$ is the 2D Fourier
transform of the 3D screened Coulomb impurity scattering matrix
element for 
a 2D electron, scattering from the 2D wave vector $\vk$ to $\vk'$,
$\theta_{\vk \vk'}$ is the scattering angle between the 2D wave
vectors $\vk$ and $\vk'$, and $\varepsilon_{\vk}=\hbar^2k^2/2m$ is the
2D electron energy. We use RPA to calculate screening in this
work \cite{dassarma1,RMP}. The two kinds of charged impurities
important for our 
consideration are the 2D modulation dopants, $N_i(z)=n_d\delta(z-d)$,
and the 3D unintentional background charge, $N_i(z)=N_{3D}$, with
$n_d$ and $N_{3D}$ denoting 2D and 3D impurity densities
respectively. To obtain the mobility, we obtain a thermal average of
$\tau^{-1}(\varepsilon_{\vk})$ over the finite temperature Fermi
distribution function following the Boltzmann transport theory
prescription \cite{dassarma1,RMP}.

For calculating the acoustic phonon scattering contribution to 2D
carrier transport, we follow the Boltzmann theory developed by
Kawamura and Das Sarma \cite{kawamura,kawamura2}, considering
scattering through 
both the deformation potential coupling and the piezoelectric coupling
taking into account only the GaAs acoustic phonons. We do not show
here the theoretical details, referring instead to the original work
\cite{kawamura,kawamura2} for the standard technical details. We include
screening effects in our phonon scattering calculations.

Before presenting our calculated mobility results, we mention that our
quantitative results 
depend on the parameters $d$, $n_d$, $N_{3D}$, and $D$,
characterizing respectively the remote modulation dopant scattering
($d$, $n_d$), the background unintentional dopant scattering
($N_{3D}$), and the GaAs 
deformation potential coupling constant ($D$) for the acoustic phonon
scattering. One expects $n_d=n$, where $n$ is the 2D carrier
density, in  the  modulation doped structures, but for the sake of
generality, we will assume $n_d$ to be an independent variable,
which is allowed in the presence of an external gate or compensating
impurity charges. The precise value of the deformation potential
coupling constant $D$ to be used in 2D GaAs carrier transport
consideration has been controversial as discussed in refs.
\onlinecite{kawamura} and \onlinecite{kawamura2}, 
with values of $D$ differing by a factor of two ($D=7-14$eV) being
quoted in the literature. Since $\mu \sim |D|^{-2}$, this is {\it not}
a trivial matter as the calculated mobility will be uncertain by a
factor of four depending on the precise choice of $D$. We follow
earlier transport theories, most notably \cite{kawamura} and choose
$D=10$eV as the most suitable value for the GaAs acoustic phonon
deformation coupling constant. Other values of $D$ would reflect in
an appropriate $(D/10)^2$ rescaling of our calculated phonon
mobilities.

\begin{figure}
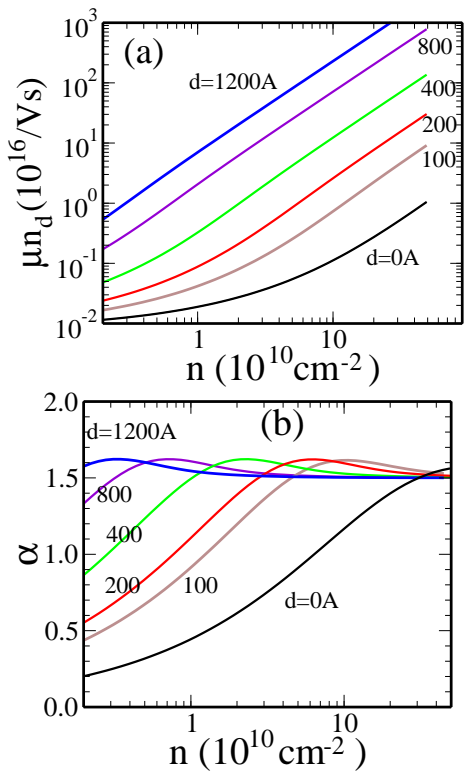

\epsfysize=2.in
\centerline{\epsffile{fig1a.eps}}
\epsfysize=2.in
\centerline{\epsffile{fig1b.eps}}
\caption{(Color online)
(a) Calculated mobility times 2D impurity density ($\mu n_{d}$)
as a function of electron density for various values of $d$, and
(b) exponents ($\alpha$) of mobility in $\mu\propto
n^{\alpha}$.}
\end{figure}

All our results are for quantum wells of width 300 \AA \; which are
symmetrically delta-doped on both sides with a modulation doping layer
separation of $d$.
We use 300 \AA \; wide symmetrically doped
AlGaAs-GaAs-AlGaAs quantum wells for our calculations because recent
ultra-high fabricated high-density 2D systems have mostly been
fabricated in quantum well device (since higher carrier densities can
be achieved in such symmetrically modulation doped devices leading to
higher carrier mobilities). All the qualitative trends we obtain
remain valid for both modulation-doped AlGaAs-GaAs heterostructure and
quantum well systems with only minor quantitative variations arising
from variations in system widths. Subband wavefunction effects
\cite{dassarma1,RMP} are
explicitly included in our theoretical calculations of various
scattering matrix elements and the screening function, making our
theory quantitatively reliable. We neglect all inter-subband scattering,
nonlinear screening, and localization effects.

In Fig. 1(a)  we show our $T=0$ mobility results for remote dopant
scattering as a function of the electron density 
%$n$ varying from $n=5\times
%10^{9}cm^{-2} - 5\times 10^{11}cm^{-2}$ 
for various values of the
modulation doping separation distance $d=0-1200$\AA \; (note that
$d=0$ implies that the charged impurities are at the GaAsAs-GaAs
interface). Since $\mu \propto n_d^{-1}$, we show in Fig.\;1 the
product 
$\mu n_{d}$ as a function of $n$ for various values of $d$. Our
results at the lowest and the highest carrier densities are not
reliable due to 
our neglect of localization and intersubband scattering effects,
respectively, but the results should be valid in the
$n=10^{10}-3\times 10^{11}$ cm$^{-2}$ range without any problem for
high-quality samples.

\begin{figure}
\epsfysize=2.2in
\centerline{\epsffile{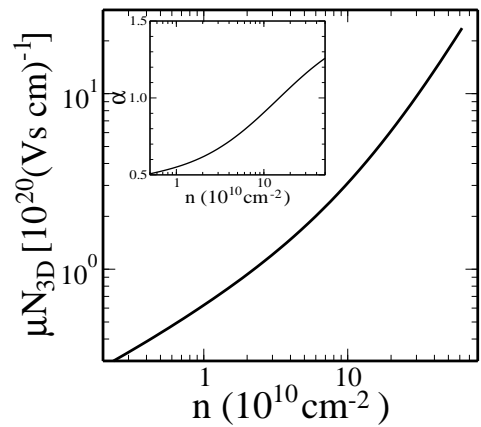}}
\caption{
(a) Calculated mobility times 3D impurity density ($\mu \times
N_{3D}$) as a function of electron density. In inset the exponent
$\alpha$ is shown.
}
\end{figure}

Assuming a high value of $n_{d}=10^{11}$ cm$^{-2}$, we see that the
maximum intrinsic mobility at $n=10^{11}$ cm$^{-2}$ increases from
$3\times 10^5$ cm$^2$/Vs for $d=100$\AA \; to the very large number of
$30\times 10^6$ cm$^2$/Vs for $d=1200$\AA. Since the maximum achieved
low-temperature 2D mobility so far has been around $36\times
10^6$ cm$^2$/Vs for $d \agt 1000$\AA, we conclude that the remote dopants
are not the primary mobility limiting mechanism although they may be
contributing about 30\% of the resistive scattering. In Fig. 1(b) we show
that approximate density exponent ($\alpha$) of mobility, which has
often been discussed in the literature, by writing $\mu \sim
n^{\alpha}$, i.e. $\alpha \equiv d\ln \mu/d\ln n$. We note that except
for $d=0$, $\alpha>1$ for remote impurity scattering, and $\alpha \approx
1.5$ for larger values of $d$. This is simply a reflection of the fact
that the remote scatterers are rather ineffectively screened by the 2D
carriers, leading to $\alpha \approx 3/2$ which corresponds to the
unscreened limit. We conclude from Fig. 1 that, if all other
scattering mechanisms can be eliminated (most importantly, the
unintentional background impurities), then at low temperatures, the 2D
mobility could easily exceed $100\times 10^6$ cm$^2$/Vs at $n=3\times
10^{11}$ cm$^{-2}$ provided $d\agt 1000$\AA.
Note that for large $d$, where $k_F d \gg 1$ with $k_F \sim \sqrt{n}$
being the 2D Fermi wave vector, $\mu \sim d^3$, and therefore, as a
matter of principle, remote dopant scattering can be reduced
indefinitely simply by increasing $d$ although in practice this is
difficult since inducing carriers in the 2DEG becomes difficult for
large $d$.

In Fig. 2, we show our calculated mobility due to scattering by
unintentional background charged impurities using the same quantum
well parameters, but now with a three-dimensional uniform random
background distribution of charged impurities of concentration
$N_{3D}$ per unit volume which reside everywhere (both in GaAs and in
AlGaAs). Again, $\mu \propto N_{3D}^{-1}$, and
therefore we show in Fig. 2 our calculated $\mu N_{3D}$ as a function
of the carrier density $n$ at $T=0$. For very clean and pure materials
as typically used in the very best MBE growth laboratories, $N_{3D}$
is extremely small, and cannot be directly measured by any
spectroscopic tools (e.g. DLTS). Assuming $N_{3D}=4\times 10^{13}$
cm$^{-3}$, 
which is a low estimate, we get $\mu \approx 30\times 10^{6}$
cm$^{2}$/Vs 
at $n=3\times 10^{11}$ cm$^{-2}$. It therefore appears that the current
limit to the achievable 2D mobilities in GaAs systems is set entirely
by the background unintentional charged impurities, and reducing their
level to $10^{12}$ cm$^{-3}$ or below should raise the 2D mobility to
$100\times 10^6$ cm$^2$/Vs or above at low temperatures.
To further reinforce the relevance of background scattering, we show as
an inset to Fig. 2 the exponent $\alpha$ (i.e. $\mu \sim n^{\alpha}$)
for background scattering by calculating $\alpha = d\ln \mu/d \ln
n$. The value of $\alpha \alt 1$ agrees well with the
corresponding experimentally measured exponent. Note that the exponent
$\alpha$ decreases as the width of quantum well decreases and the
measured exponent in the ultrahigh mobility system is considerably
below $\alpha \approx 1.5$ implied by remote impurity scattering
[Fig. 1(a)].

\begin{figure}
\epsfysize=2.4in
\centerline{\epsffile{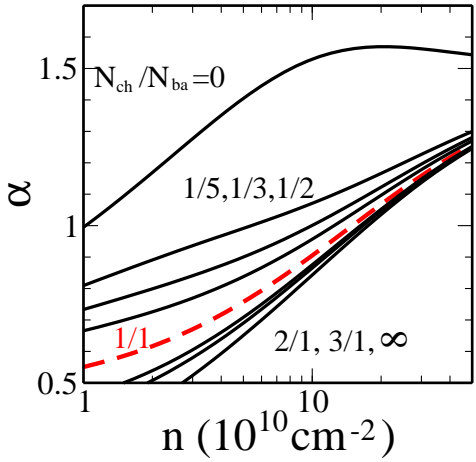}}
\caption{
The mobility exponent $\alpha$ in $\mu\propto n^{\alpha}$ for
different $N_{ch}/N_{ba}=0$, 1/5, 1/3, 1/2, 1, 2, 3, $\infty$ (from
top to bottom). Here $N_{ch}$ ($N_{ba}$)
indicates the channel (barrier) 3D impurity density. The dashed line
represents $N_{ch}=N_{ba}$.
}
\end{figure}

\begin{figure}
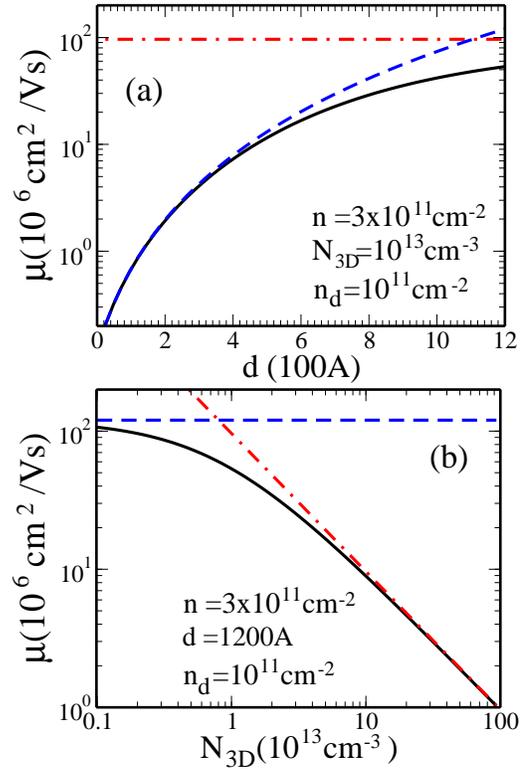

\epsfysize=2.in
\centerline{\epsffile{fig4a.eps}}
\epsfysize=2.in
\centerline{\epsffile{fig4b.eps}}
\caption{(Color online)
(a) Calculated mobility as a function of $\delta$-layer distance $d$
for fixed values of the densities $n=3\times 10^{11}cm^{-2}$,
$n_d=10^{11}cm^{-2}$, and $N_{3D}=10^{13}cm^{-3}$ 
and (b) as a function of 3D impurity $N_{3D}$ for $n=3\times
10^{11}cm^{-2}$, $n_d=10^{11}cm^{-2}$, and $d=1200$\AA.
The dashed 
(dot-dashed) curves in (a) and (b) show remote (background) charged
impurity contribution. 
}
\end{figure}

Since the current low temperature ($\alt 1K$) 2D mobility limit in
high-quality GaAs structures is set by the Coulomb scattering by
charged impurities, mostly by the unintentional background impurities
and partially by the remote dopants in the modulation layer, we show
in Figs. 3 and 4 some more details on the theoretical mobility arising
from the screened Coulomb scattering. Noting that the quantitative
details of the background unintentional doping are simply unknown
(since their concentration is far too low to be studied
spectroscopically), we show in Fig. 3, the calculated mobility
exponent $\alpha$ (i.e. $\mu \sim n^{\alpha}$) for different
combinations of the barrier (i.e. AlGaAs) and the channel (i.e. GaAs)
background impurity scattering leaving out the remote dopant
scattering. The dashed curve with $N_{ch}/N_{ba} =1$ in Fig. 3
indicates the results shown in Fig. 2 where the 3D random charged
impurities are distributed uniformly throughout the background
(i.e. both in the GaAs channel and in the AlGaAs barrier). As
expected, the exponent $\alpha$ decreases as the relative background
scattering from the GaAs layer increases with respect to that from the
AlGaAs barrier region. This is a direct effect of screening by the 2D
carriers themselves which is, of course, stronger (weaker) for the
channel (barrier) impurities. For $n\approx 10^{11}$ cm$^{-2}$, the
exponent $\alpha \sim 0.7$ is in excellent agreement with experimental
results in the highest mobility variable density gated structures
\cite{lilly} where the 2D mobility has been measured as a function of
carrier density. It is noteworthy that in Fig. 3, $\alpha$ increases
with increasing carrier density, reflecting the well-known
\cite{dassarma1,RMP} peculiarity of 2D screening that it becomes
weaker (stronger) with increasing (decreasing) carrier density since
the relevant dimensionless screening parameter $q_{TF}/2k_F$, where
$q_{TF}$ ($k_F$) are 2D Thomas-Fermi screening wave vector (Fermi wave
vector) goes as $q_{TF}/2k_F \sim n^{-1/2}$ by virtue of $q_{TF}$ being
density independent in 2D. Thus, for very large density, $q_{TF}/2k_F
\ll 1$, we recover the unscreened result $\alpha \sim 3/2$. At 
high density ($n \gg 10^{11}$ cm$^{-2}$), it is possible that interface
roughness scattering (not included in our theory) could start
contributing as the 2D electrons are pushed close to the GaAs-AlGaAs
interface by the self-consistent confining potential, leading to
$\alpha \sim 0$ since the interface roughness is a short-range
scattering potential with very weak density dependence.

\begin{figure}
\epsfysize=2.2in
\centerline{\epsffile{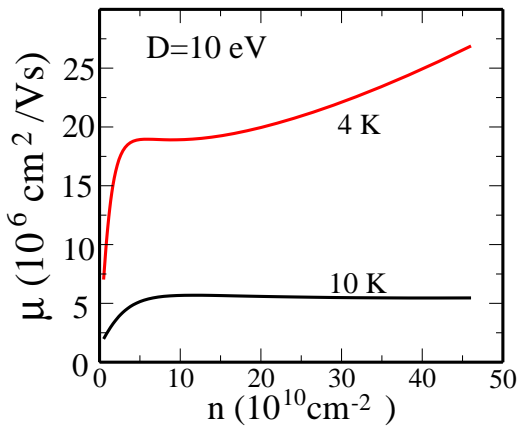}}
\caption{
(a) Calculated acoustic phonon-limited mobility 
as a function of density for two different densities. $D=10$eV is used
in this calculation.
}
\end{figure}

\begin{figure}
\epsfysize=2.5in
\centerline{\epsffile{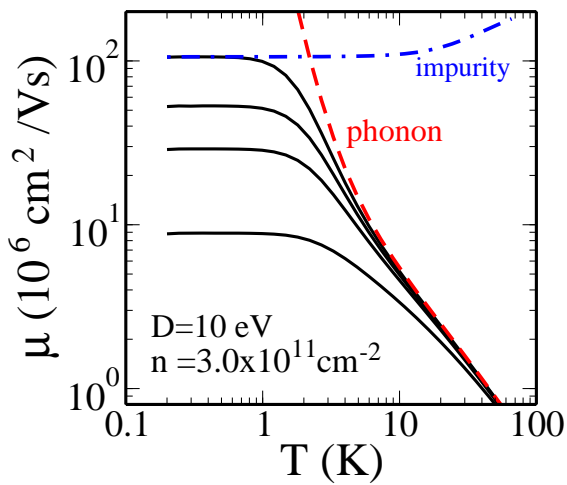}}
\caption{
Calculated total mobility (black lines) as a function of
temperature for a fixed electron density $n=3 \times 10^{11}$ cm$^{-2}$
for different 3D impurities, $N_{3D}=0.1$, 1.0, 2.5, 10$\times
10^{13}$ cm$^{-3}$ (top to bottom) with $d=1200$ \AA \; and $n_{d} =
10^{11}$ cm$^{-2}$. The dashed 
(dot-dashed) line shows phonon (charged impurity for $N_{3D}=0.1\times
10^{13}$ cm$^{-3}$) only contribution.  
}
\label{fig4}
\end{figure}

In Fig. 4, we show the combined effects of remote and background
Coulomb scattering (i.e. a combination of Figs. 1 and 2) on the 2D
mobility at $T=0$. In Fig. 4(a), we show our calculated mobility as a
function of the modulation separation $d$ for fixed values of the
densities $n$, $n_d$, and $N_{3D}$. It is clear that for $n_d =
10^{11}$ cm$^{-2}$ and $N_{3D}=10^{13}$ cm$^{-3}$, $\mu$ reaches a maximum
of around $50 \times 10^6$ cm$^2$/Vs for $d \approx 1200$\AA \; since
the background impurities become dominant for large $d$. Any further
enhancement of mobility would necessitate decreasing $N_{3D}$, which
is shown in Fig. 4(b), where we plot $\mu$ as a function of $N_{3D}$
for fixed values of $d$, $n_d$, and $n$. A 2D mobility of $100 \times
10^6$ cm$^2$/Vs (the `100 millions' in the title of our paper) can be
achieved if $N_{3D}$ can be brought down to $10^{12}$ cm$^{-2}$, an
extremely small, but not an impossible, number. Results shown in
Fig. 4 explicitly demonstrate how depending on the system parameters
$n$, $d$, $n_d$, and $N_{3D}$, the actual mobility can be enhanced
either by increasing $d$ or by decreasing $N_{3D}$, but extremely
large 2D mobility would necessarily require large $d$ ($\agt 1000$\AA)
and small $N_{3D}$ ($\alt 10^{13}$ cm$^{-3}$). We show our intrinsic
phonon scattering results next, emphasizing that for $T \alt 1$K, the
results of Figs. 1--4 shown above remain unaffected by phonon
scattering.

In Fig. 5 we consider the mobility due to acoustic phonon scattering,
which is shown  as a function of carrier density for $T=4$K
and 10K. For lower temperatures, $\mu_{ph}$ increases by a large
factor ($\mu \sim T^{-7}$ for deformation potential scattering and
$\mu \sim T^{-5}$ for piezoelectric scattering) since one is in the
Bloch-Gr\"uniesen 
range where phonon accupancy is suppressed exponentially
\cite{stormer_ph}. The  
important point of Fig. 5 is that at $T=10$K (or even at 
4K), the mobility at $n=3\times 10^{11}$ cm$^{-2}$ is totally dominated
by phonons and becomes $5\times 10^6$ 
cm$^2$/Vs ($22\times 10^6$ cm$^2$/Vs), which is much lower than that for
charged impurity scattering. We therefore conclude that it will be
impossible to raise 2D mobility above $22 \times 10^6$ ($5\times
10^6$) cm$^2$/Vs at $T=10$ (4)K since acoustic phonon scattering sets
the intrinsic limit at these higher temperatures. At 1K or below,
however, acoustic phonon scattering is completely suppressed, and the
mobility is limited by background charged impurity scattering and, to
a lesser degree, by the remote dopant scattering.
For example, at $T=2K$ the acoustic phonon limited mobility exceeds
100 millions.

Finally, as a conclusion, we combine the acoustic phonon and
background impurity scattering in Fig. 6, by showing how progressive
reduction of the unintentional background charge would raise the
mobility toward 100 million cm$^2$/Vs provided $d \agt 1000$\AA. We
estimate that a background 
impurity density of $10^{12}$ cm$^{-3}$ 
would produce a 2D mobility of $10^8$
cm$^2$/Vs at $T=1K$ and $n=3\times 10^{11}$ cm$^{-2}$ for $d=1200$\AA \;
and $n_d=10^{11}$ cm$^{-2}$.

This work is supported by Microsoft Q project, DOE Sandia National
Lab, and LPS-NSA-CMTC.

%%%%%%%%%%%%%%%%%%%%%%%%%%%%%%%%%%%%%%%%%%

\end{document}